\documentclass[pra,twocolumn,showpacs]{revtex4}
\usepackage{graphicx}
\usepackage{amssymb}
\newcommand{\beq}{\begin{equation}}
\newcommand{\enq}{\end{equation}}
\newcommand{\bea}{\begin{eqnarray}}
\newcommand{\ena}{\end{eqnarray}}
\newcommand{\rr}{\mathbf{r}}
\begin{document}

\title{Correlations and superfluidity of a one-dimensional Bose gas 
in a quasiperiodic potential}
\author{Alberto Cetoli}
\affiliation{Department of Physics, Ume{\aa} University, \
  SE-90187 Ume{\aa}, Sweden}
\author{Emil Lundh}
\affiliation{Department of Physics, Ume{\aa} University, \
  SE-90187 Ume{\aa}, Sweden}
\begin{abstract}
  We consider the correlations and superfluid properties of a Bose gas
  in an external potential.  Using a Bogoliubov scheme, we obtain
  expressions for the correlation function and the superfluid density
  in an arbitrary
  external potential. 
  These expressions are 
  applied to a one-dimensional system at 
  zero temperature subject to a quasiperiodic modulation.  The
  critical parameters for the Bose glass transition are obtained using
  two different criteria and the results are compared.  The Lifshits
  glass is seen to be the limiting 
  case for vanishing interactions. 
\end{abstract}
\maketitle

\section{Introduction} \label{sec:intro} 

Research on the problem of a Bose gas in a disordered potential gained
momentum with the realization of well-controlled disordered potentials
for ultracold bosonic atoms, which inspired a surge in theoretical and
experimental activity
\cite{fallani2007,astrakharchik2002,louis2007,lugan2007,roux2008,deng2008,fontanesi2009},
and the subsequent observation of Anderson localization by two groups
\cite{roati2008,billy2008}. The quest is now on to better understand
the effect of disorder on an interacting quantum system. 

The phase diagram of an interacting Bose gas subject to a disordered
potential was outlined in, e.g., Refs.\
\cite{fisher1989,rapsch1999,astrakharchik2002,louis2007,lugan2007,
  roux2008,deng2008,fontanesi2009}. While the exact picture of the
phase diagram depends on the employed potential, the qualitative
behavior appears to be the same for all kinds of disorder
realizations.  In absence of disorder, the gas is in a superfluid
phase, which in a two- or three-dimensional Bose system at zero
temperature is identical to a Bose-Einstein condensate (BEC), and in
one dimension is a quasi-condensate with suppressed density
fluctuations and algebraically decaying correlations. Upon raising the
disordered potential, the system can enter a new quantum phase: the
Bose glass state, also called a fragmented Bose-Einstein condensate.
The Bose glass state lacks long-range phase coherence, having
exponentially decaying correlations and zero superfluid density, but
it is compressible. In the weakly interacting limit, it has been
conjectured that the Bose glass goes over into a Lifshits (or
Anderson) glass \cite{lugan2007,deng2008}, where the density profile
is that of a superposition of the lowest-lying single-particle states,
which are exponentially localized. Finally, a gas of noninteracting
bosons will have all the particles occupying the lowest of the
single-particle states.

Experimentally, two basic types of disordered potential for ultracold
atoms have been realized, both by optical means; speckle and
quasiperiodic. The first type of disorder uses laser speckle, the
two-dimensional diffraction pattern of a laser beam passing through a
roughened plate. Speckle potentials for ultracold atoms were first
used in Ref.  \cite{clement2005}. A quasiperiodic potential in one
dimension is created simply by superposing two standing waves with
different wavelength; if the wavelengths differ by an irrational
factor -- the most popular being the golden ratio -- the resulting
intensity is a quasiperiodic function of the coordinate
\cite{fallani2007}.  In higher dimensions, quasiperiodic potentials
can be created as the interference pattern of a number of beams
meeting at judiciously chosen angles \cite{guidoni1999}. 

The speckle potential is intended to achieve the effect of a
one-dimensional random pattern. This situation has been studied
extensively in the literature. For a weakly interacting (repulsive)
system, Albanese and Fr\"olich \cite{albanese1988} showed that the
Hamiltonian admits exponentially localized solution in the presence of
arbitrarily small disorder. On the other hand, an interacting 1D Bose
gas acts as a Luttinger Liquid. Using this approach, Giamarchi and
Schultz \cite{giamarchi1988} showed that for a strong enough
interaction the system appears again in a localized phase. In order to
match with the known results in the weakly interacting limit, they
speculated the existence of two localized phases. Later on, the
existence of those two sides of the glass transition has been shown
numerically in Ref.  \cite{Scalettar1991} and \cite{Prokofev1998}.

In the quasiperiodic case the situation is less clear. While it is
established that - for no interaction - the system is exponentially
localized only when the external potential is strong enough
\cite{eliasson1992}, in the presence of a repulsive interaction the
exact behavior of the localization is still not known, and only recent
results \cite{adhikari2008} show that a quasiperiodic potential can
inhibit the diffusion of a wavepacket. In this kind of potential the
phase diagram has been studied by Ref. \cite{roux2008,deng2008}. These
works are done under the assumption that the secondary lattice is
small compared to the primary one, and the system can be mapped into a
Bose Hubbard (BH) model. However, very little is known when the
secondary lattice strength is comparable to the primary one.  In this
case the BH model cannot be used, and one has to address the problem
in the continuum. Moreover, if there exists an underlying periodic
potential to which the disorder is added, there is also a Mott
insulating phase for strong interactions if the filling is integer.
The latter feature will not be discussed much in the present paper
because it is not correctly described by the approximation that we
intend to use.

We see from the above discussion that the correlation properties is a
key to understanding the behavior of the Bose gas; the superfluid
properties is another. Both of these can be calculated in the
Bogoliubov scheme, where the Gross-Pitaevskii equation gives the
(quasi-) condensate density and superfluid velocity and the Bogoliubov
equations yield the excitation spectrum and corrections to the
density. The Bose glass transition in a Gaussian correlated potential 
was explored recently for strong interparticle potentials by
Fontanesi {\it et al.} in Ref.\ \cite{fontanesi2009}. 
The Bogoliubov scheme is usually derived starting from the
assumption that the system is Bose-Einstein condensed and there exists
a condensate wavefunction $\Phi(\rr)$, defined either as the
expectation value of the bosonic field operator or as the wavefunction
for the single-particle mode that is occupied by a macroscopic
fraction of the particles. Clearly, this assumption would appear to
preclude the description of both quasicondensate and Bose glass, but,
as we shall see, it turns out that the Bogoliubov scheme can in fact
describe those states as well.  The explanation is that BEC is
convenient, but not a necessary requirement, for deriving the
Bogoliubov equations; it suffices to make the weaker assumption
that the quantum fluctuations in the density and gradient of the phase
are small. Several papers have presented different sketches of the
derivation of Bogoliubov theory for a quasicondensate
\cite{shevchenko1992,wu1996,ho1999,alkhawaja2002,mora2003,xia2005}.
In this article we put up what we believe to be a complete, simple and
consistent derivation of the Bogolioubov equations in a Bose gas using
a minimum of assumptions, in order to be able to analyze the
quasicondensate - Bose glass transition in one dimension.  Using the
same formalism we derive an expression for the superfluid density in
order to better understand the transition.

In this work we consider the experimentally relevant case of a 1D Bose
gas in a quasiperiodic potential and examine a wide parameter regime,
allowing for a discussion of the Bose glass transition and the
conjectured Lifshits glass. It is important to stress that the
Bogoliubov approximation has to be used with some care in
one-dimensional systems. This problem has been treated in detail for a
uniform gas by Lieb and Lininger \cite{leib_lininger1963}, where it is
shown that the Bogoliubov perturbation theory agrees with the exact
answer for $g/n < \hbar^2/2\,m$. Moreover, our numerical scheme is
consistent only if the fluctuation in the phase and in the density are
small, and this clearly cannot be the case upon increasing
indefinitely the interaction within the particles. For this reason,
due to the intrinsic limit of the Bogoliubov approach, we expect our
results to be reliable only for weak interaction and high density.
Within these limits, the application of the Bogoliubov recipe has been
proven to be theoretically sound, and it is known to give the same
correlation function of the Luttinger liquid theory, as shown in Ref.
\cite{andersen2002}. We wish to compare the different predictions for
the phase transition obtained from analyzing the behavior of the
correlation function and of the superfluid density.

As we shall see, our analysis suggests the existence of a glassy phase
for small interparticle interaction. The plot of this phase appears as
a series of peaks with little overlap. According to the language used
in Ref \cite{lugan2007}, this phase is called ``fragmented BEC'', or
Bose glass; in the same reference, the Anderson glass appears when the
overlap between the peaks is negligible. We maintain these definitions
through our article.

This paper is organized as follows. In Section \ref{sec:correlation} 
we present a derivation of Bogoliubov theory and in particular 
an expression for a correlation function. 
In Sec.\ \ref{sec:superfluid}, an expression for the superfluid density 
is derived. 
Sec.\ \ref{sec:boseglass} presents numerical results for the 
superfluid-Bose glass phase transition in a 1D Bose gas. 
Finally, in Sec.\ \ref{sec:conclusion} we summarize and conclude.

\onecolumngrid

\section{Correlation function} \label{sec:correlation} 

Inspired by the works of Ho and Ma \cite{ho1999}, and Xia and Silbey 
\cite{xia2005}, we
use a path integral formalism, considering a system of bosons
described by the Euclidean action
\begin{equation}
\mathcal{S}[\psi^*,\psi]
= \int d\tau\,d\mathbf{r}
       \,
       \psi^*(\mathbf{r},\tau)
       \,
       \left(
         -\hbar\,\partial_\tau
         + \frac{\hbar^2}{2\,m}\nabla^2
         - U(\mathbf{r})
         + \mu
         -\frac{g}{2} \, |\psi(\mathbf{r},\tau)|^2
       \right)
       \, \psi(\mathbf{r},\tau)
\, .
\end{equation}
where $\psi$ is the scalar boson field. Assuming that the quantum
fluctuations of density and gradient of the phase are small we expand $\psi$
as
\begin{eqnarray} \label{PsiExp}
\psi(\mathbf{r},\tau) 
&=& e^{i\,\theta(\mathbf{r},\tau)}
    \,\sqrt{n_0(\mathbf{r})+\delta n(\mathbf{r},\tau)}
\nonumber \\
&\approx& e^{i\,\theta(\mathbf{r},\tau)}
    \,\sqrt{n_0(\mathbf{r})} 
    \left( 1 + \frac{1}{2}\frac{\delta n(\mathbf{r},\tau)}{n_0(\mathbf{r})}
             -\frac{1}{8}\frac{\delta n^2(\mathbf{r},\tau)}{n_0(\mathbf{r})^2}
    \right)
\nonumber \\
&=& \psi_0(\mathbf{r},\tau) + \delta\psi(\mathbf{r},\tau)
\, .
\end{eqnarray}
The equation of motion for $\psi_0$ is found by means of the
variational principle $\delta \mathcal{S}/\delta n_0^*=0$, and the
result is the Gross-Pitaevskii equation for $n_0$ \cite{pethick2008book}, 
\begin{equation}
-\frac{\hbar^2}{2\,m}\,\nabla^2\sqrt{n_0(\mathbf{r})}
+U(\mathbf{r})\,\sqrt{n_0(\mathbf{r})}
+g\,n_0(\mathbf{r})^{\frac{3}{2}}
=
\mu\,\sqrt{n_0(\mathbf{r})}\,.
\end{equation}
This amounts to treating $n_0(\mathbf{r})$ as a scalar field
representing the (quasi-) condensate density, and $\delta n$, $\theta$
as perturbations. In a three-dimensional system, and in two dimensions 
at zero temperature, 
the condensate wavefunction $\Phi$ would in the absence of currents 
be equal to the square root of 
the classical density, $\sqrt{n_0}$. 
In principle one could describe a state with a
current by assuming a slowly varying phase $\theta_0$ that is also
treated classically, but we refrain from doing so for convenience.

Ignoring terms of order higher than $\theta^2$ and $\delta n^2$ the
action becomes
\begin{equation}
\mathcal{S}= \mathcal{S}_0 + \mathcal{S}_1 + \mathcal{S}_2
\, .
\end{equation}
Here, $\mathcal{S}_0$ contains only the classical density $n_0$ and 
is minimized by the equation of motion for
$n_0$, $\mathcal{S}_1$ must vanish for $n_0$ to be a stationary
solution, while $\mathcal{S}_2$ is
\begin{equation}
\mathcal{S}_2= 
\frac{1}{2}
\int
\,d\tau\,d\mathbf{r}
\,
\left( 
  \begin{array}{c}
    \frac{\delta n(\mathbf{r},\tau)}{\sqrt{2\,n_0(\mathbf{r})}} \\
    i\,\sqrt{2\,n_0(\mathbf{r})}\,\theta(\mathbf{r},\tau)
  \end{array}
\right)^*
S
\,
\left( 
  \begin{array}{c}
    \frac{\delta n(\mathbf{r},\tau)}{\sqrt{2\,n_0(\mathbf{r})}} \\
    i\,\sqrt{2\,n_0(\mathbf{r})}\,\theta(\mathbf{r},\tau)
  \end{array}
\right) 
\, ,
\end{equation}
with 
\begin{eqnarray} \label{S2}
S 
&=&
\left( 
  \begin{array}{cc}
    -\frac{\hbar^2\,\nabla^2}{2\,m} + U + 3\,g\,n_0 - \mu
       & 
    -\hbar\,\partial_\tau \\
    -\hbar\,\partial_\tau 
       & 
    -\frac{\hbar^2\,\nabla^2}{2\,m} + U + g\,n_0 - \mu
  \end{array}
\right) \nonumber \\ 
&=&
-\hbar \partial_\tau \, \sigma_1 
 + 
\left( 
  \begin{array}{cc}
    H_3 & 0\\
    0 & H_1\\
  \end{array}
\right) 
\, ,
\end{eqnarray}
where we have implicitly defined the scalar operators $H_3$ and $H_1$.
In order to find the correlators between $\delta n$ and $\theta$ we
need to find the function $G$ that inverts $S$, 
\begin{equation} \label{toInvert}
\left[
-\hbar \partial_\tau \,\sigma_1 
 + 
\left( 
  \begin{array}{cc}
    H_3 & 0\\
    0 & H_1\\
  \end{array}
\right) 
\right]
\,
G(\mathbf{r},\tau,\mathbf{r}',\tau')
= \delta(\mathbf{r}-\mathbf{r}')
\,.
\end{equation}
Of course, what we have derived so far is mathematically identical to 
the well-known Bogoliubov theory, as we now show. 
Introducing the transformation $T$ as 
\begin{eqnarray}
T
&=&
\frac{1}{\sqrt{2}}
\,
\left( 
  \begin{array}{cc}
    1 &  1 \\
    1 & -1
  \end{array}
\right)
\,,
\end{eqnarray}
we obtain
\begin{eqnarray}
T \, \mathcal{L} \, T^{-1}
 = 
\left( 
  \begin{array}{cc}
     0     & H_1\\
     H_3   & 0
  \end{array}
\right) 
\, ,
\end{eqnarray}
with 
\begin{eqnarray}
\mathcal{L}
 = 
\left( 
\begin{array}{cc}
-\frac{\hbar^2}{2\,m} \nabla^2
 + U 
 + 2\,g\,n_0 
 - \mu\        & n_0\,g\\
 - n_0\,g      & \frac{\hbar^2}{2\,m} \nabla^2
                 - U 
                 - 2\,g\,n_0 
                 + \mu
\end{array}
\right) \,.
\end{eqnarray}
The diagonalization of $\mathcal{L}$ leads to the Bogoliubov 
equations for the Bogoliubov 
amplitudes $u_j(\mathbf{r})$ and $v_j(\mathbf{r})$,
\begin{eqnarray} \label{BdG}
\mathcal{L}
\,       \left( \begin{array}{c} u_j\\ v_j \end{array} \right)
= E_j \, \left( \begin{array}{c} u_j\\ v_j \end{array} \right)
\,.
\end{eqnarray}
Moreover, the Green function for the action $S=i\hbar\,\partial_t +
\mathcal{L}$ is already known \cite{Lundh2002b}, and equal to
\begin{eqnarray}
G(\mathbf{r},\mathbf{r}',\omega)
=
-\sum_{j\ne0}
    \hbar
    \left[
    \frac{1}{\hbar\omega - E_j}\,
    \left( 
      \begin{array}{c}
        u_j \\
        v_j \\
      \end{array}
    \right) 
    \,
    \left( 
      \begin{array}{c}
        u_j' \\
        v_j' \\
      \end{array}
    \right)^{\dagger}
    - 
    \frac{1}{\hbar\omega + E_j}
    \left( 
      \begin{array}{c}
        v_j^* \\
        u_j^* \\
      \end{array}
    \right) 
    \,
    \left( 
      \begin{array}{c}
        {v_j'}^* \\
        {u_j'}^* \\
      \end{array}
    \right)^\dagger
    \right]\,.
\end{eqnarray}
Defining $\chi_j$ as
\begin{eqnarray} \label{Chi}
\chi_j
=
\left( 
  \begin{array}{c}
    \chi_j^1 \\
    \chi_j^2 
  \end{array}
\right)
=
T
\,
\left( 
  \begin{array}{c}
    u_j \\
    v_j 
  \end{array}
\right)
=
\left( 
  \begin{array}{c}
    \frac{\delta n_j}{\sqrt{2}\,\sqrt{n_0}} \\
    i\,\sqrt{2\,n_0}\,\theta_j
  \end{array}
\right) 
\, ,
\end{eqnarray}
and correspondingly 
\begin{eqnarray} \label{ChiTilde}
\tilde{\chi}_j
&=&
T
\,
\left( 
  \begin{array}{c}
    v_j^* \\
    u_j^* 
  \end{array}
\right)
\, ,
\end{eqnarray}
we can apply the transformation $T$ to Eq.\ (\ref{toInvert}) and eventually
find
\begin{eqnarray} \label{Gphipsi}
G(\mathbf{r},\tau,\mathbf{r}',\tau^+)
&=& \sum_{\omega_n} \frac{e^{i\,\omega_n\,\eta}}{\beta}
    G(\mathbf{r},\mathbf{r},\omega_n) \nonumber \\
&=& -\sum_{j\ne0} \sum_{\omega_n} \frac{e^{i\,\omega_n\,\eta}}{\beta}
    \left[
    \frac{1}{i\,\hbar\omega_n - E_j}\,\chi_j\,{\chi'}_j^{\dagger}
    - 
    \frac{1}{i\,\hbar\omega_n + E_j}
    \tilde{\chi}_j\,\tilde{\chi'}_j^{\dagger} 
    \right] \nonumber \\
&=& \sum_{j\ne0} \chi_j\,{\chi'}_j^{\dagger}\,N(E_j)
           + \tilde{\chi}_j\,\tilde{\chi'}_j^{\dagger}\,(N(E_j) + 1) 
    \nonumber \\
&=&
\left( 
  \begin{array}{cc}
     \frac{1}{2\,\sqrt{n_0\,n_0'}} \langle \delta n \, \delta n' \rangle 
         & - \sqrt{\frac{n_0'}{n_0}} \langle \delta n \, i\,\theta' \rangle \\
     \sqrt{\frac{n_0}{n_0'}} \langle i\,\theta \, \delta n \rangle
         & 2\, \sqrt{n_0\,n_0'} \langle \theta \,\theta' \rangle
  \end{array}
\right)
\, ,
\end{eqnarray}
with $\omega_n$ being the Matsubara frequencies, $\eta$ a positive 
infinitesimal, and
$N(E_j)=1/(e^{\beta\,E_j}-1)$ is the Bose-Einstein distribution function. 
For convenience we have defined the notation
\begin{eqnarray}
\theta
&=& 
\theta(\mathbf{r},\tau)
\nonumber \\
\theta'
&=& 
\theta(\mathbf{r}',\tau^+)
\nonumber \\
\delta n
&=& 
\delta n(\mathbf{r},\tau)
\nonumber \\
\delta n'
&=& 
\delta n(\mathbf{r}',\tau^+).
\end{eqnarray}
Furthermore let us define 
\beq
\Delta \theta
= \theta'-\theta.
\enq
Using Eq.\ (\ref{PsiExp}) the one-body correlation function becomes
\begin{eqnarray}
\langle \psi^*(\mathbf{r})\, \psi(\mathbf{r'})\rangle
&=& \langle \sqrt{n_0+\delta n} \, e^{-i\,\theta}\,e^{i\,\theta'} \, \sqrt{n_0'+\delta n'} \rangle
\nonumber \\
&=& \langle \sqrt{n_0+\delta n} \, e^{i\,\Delta\theta} \, \sqrt{n_0'+\delta n'} \rangle
\nonumber \\
&=& \sqrt{n_0\,n_0'} \langle e^{i\,\Delta \theta} 
                     + \frac{1}{2}\,\frac{\delta n}{n_0}\,e^{i\,\Delta\theta} 
                     + \frac{1}{2}\,e^{i\,\Delta\theta} \,\frac{\delta n'}{n_0'}
\nonumber \\
                     &-& \frac{1}{8}\,\left(\frac{\delta n}{n_0}\right)^2\,e^{i\,\Delta\theta}
                     - \frac{1}{8}\,e^{i\,\Delta\theta}\,\left(\frac{\delta n'}{n_0'}\right)^2
                     - \frac{1}{4}\,\frac{\delta n}{n_0}\,e^{i\,\Delta\theta}\,\frac{\delta n'}{n_0'}
                     \rangle
\,.
\end{eqnarray}
This expression can be evaluated using Wick's theorem. To lowest order in 
$\delta n$ and $\theta$ one finds \cite{mora2003} 
\begin{eqnarray}
\langle e^{i\,\Delta\theta}\rangle 
&=& 
e^{-\frac{1}{2}\,\langle\,(\Delta\theta)^2\rangle} 
\nonumber \\
\langle \frac{\delta n}{n_0} 
        \, e^{i\,\theta}
\rangle 
&=& 
e^{-\frac{1}{2}\,\langle(\Delta\theta)^2\rangle} 
 \,\langle \frac{\delta n}{n_0}\,i\,\Delta\theta \rangle 
\nonumber \\
\langle 
\left(\frac{\delta n}{n_0}\right)^2\,e^{i\,\Delta\theta} 
\rangle
&\approx& 
e^{-\frac{1}{2}\,\langle(\Delta\theta)^2\rangle}
\,
\langle \left(\frac{\delta n}{n_0}\right)^2 \rangle
\nonumber \\
\langle \frac{\delta n}{n_0}\,e^{i\,\Delta\theta}\,\frac{\delta n'}{n_0'} \rangle
&\approx&
e^{-\frac{1}{2}\,\langle\,(\Delta\theta)^2\rangle}\,
\langle \frac{\delta n}{n_0}\,\frac{\delta n'}{n_0'} \rangle 
\, ,
\end{eqnarray}
so that, eventually, to second order \cite{ho1999}, 
\begin{eqnarray}
\langle \psi^*(\mathbf{r})\, \psi(\mathbf{r}')\rangle
= 
\sqrt{n_0\,n_0'} \, e^{-\frac{1}{2}\,\langle(\Delta\theta)^2\rangle}
\Big[
1 
&+& \frac{1}{2} \left( 
                \langle \frac{\delta n}{n_0}\,i\,\Delta\theta \rangle   
                +
                \langle i\,\Delta\theta \, \frac{\delta n'}{n_0'} \rangle 
              \right)
\nonumber \\
&+& \frac{1}{4} \, \langle \frac{\delta n}{n_0}\,\frac{\delta n'}{n_0'} \rangle
\nonumber \\
&-& \frac{1}{8} \, \left(
                     \langle \left(\frac{\delta n}{n_0}\right)^2 \rangle
                     +
                     \langle \left(\frac{\delta n'}{n_0'}\right)^2 \rangle
                   \right)
\Big] 
\, .
\end{eqnarray}
Since both $\delta n$ and $\theta$ are small, the expression between
square brackets can be thought as a first order expansion of an
exponential. Using Eqs.\ (\ref{Gphipsi}) and (\ref{Chi}) the following
expression is obtained, 
\begin{eqnarray} \label{correlation}
\ln\,g_1(\mathbf{r},\mathbf{r}')
&=&
\ln \langle \psi^*(\mathbf{r})\, \psi(\mathbf{r}')\rangle - \ln \sqrt{n\,n'}
\nonumber \\
&=&
-\frac{1}{2}
\,
\sum_{j\ne0} \left\{ |\frac{v_j}{\sqrt{n}}-\frac{v_j'}{\sqrt{n'}}|^2 + N_j\, \left[ |\frac{u_j}{\sqrt{n}}-\frac{u_j'}{\sqrt{n'}}|^2 + |\frac{v_j}{\sqrt{n}}-\frac{v_j'}{\sqrt{n'}}|^2 \right]  \right\}
\nonumber \\
&+&
i \, \frac{1}{2}\, 
  \sum_{j\ne0} \left[ I(\mathbf{r}, \mathbf{r}') + N_j I(\mathbf{r}, \mathbf{r}') \right]
\, ,
\end{eqnarray}
where we useed the shorthand $N_j=N(E_j)$. 
The quantity $i\,I(\mathbf{r}, \mathbf{r}')$ is purely imaginary, and equal to
\begin{eqnarray} \label{imag}
i\,I(\mathbf{r}, \mathbf{r}') 
&=&  \left( \frac{v_j^*}{\sqrt{n}}\,\frac{v_j'}{\sqrt{n'}} - \frac{v_j'^*}{\sqrt{n'}}\,\frac{v_j}{\sqrt{n}} \right) 
+ \left( \frac{u_j'^*}{\sqrt{n'}}\,\frac{u_j}{\sqrt{n}} - \frac{u_j^*}{\sqrt{n}}\,\frac{u_j'}{\sqrt{n}} \right) \nonumber \\
&+& \left(\frac{u_j'^*}{\sqrt{n'}}\,\frac{v_j'}{\sqrt{n'}}-\frac{v_j'^*}{\sqrt{n'}}\,\frac{u_j'}{\sqrt{n'}} \right)
+ \left(\frac{v_j^*}{\sqrt{n}}\,\frac{u_j}{\sqrt{n}}-\frac{u_j^*}{\sqrt{n}}\,\frac{v_j}{\sqrt{n}} \right) 
\, .
\end{eqnarray}
If the excitation energies and the condensate wavefunction are real,
then the Bogoliubov amplitudes $u_j$ and $v_j$ can also be chosen as
real. The imaginary part in Eq.  (\ref{imag}) therefore vanishes,
and the resulting expression coincides with that found by Mora and
Castin using a different formalism \cite{mora2003}. 
In particular, for the uniform case the
expressions for the Bogoliubov functions are known, leading to
\begin{eqnarray}
\ln\,g_1
&=&
-\frac{1}{4}
\,
\frac{1}{n_0}
\sum_{k\ne0} \left(1-\cos\left( \frac{k}{2} \,|\mathbf{r}-\mathbf{r'}|\right)
            \right)^2 
\left[
           |v_k|^2 
           + N_k\, \left( |u_k|^2 + |v_k|^2 \right)
\right]
\, .
\end{eqnarray}

In one dimension, Eq.~(\ref{correlation}) is an expression for the
correlation in a Bose system which is ultraviolet and infrared
convergent. Even at $T=0$ the sum has a finite value without
assuming a cutoff or a modification in the interparticle potential.

\section{Superfluid density} \label{sec:superfluid} According to the
two-fluid model \cite{Pines1966, LandauVol9}, the mass density of a
quantum fluid can be divided into a superfluid part $\rho_s(\mathbf{r})$
and a normal one $\rho_n(\mathbf{r})$, the total mass current being
$J(\mathbf{r})=\rho_n(\mathbf{r})\,v_s(\mathbf{r}) +
\rho_n(\mathbf{r})\,v_n(\mathbf{r})$. While the superfluid density is
in general
different from the condensate density, the superfluid velocity
\emph{is} the condensate velocity. In particular, upon imposing a
phase twist on the condensate wavefunction, the superfluid part will
be proportional to the additional kinetic energy.

It is in this sense that the superfluid density can be defined as a
response to a twist of the order parameter \cite{Fisher1973,
  Weichman1988}, by means of rewriting Eq.~(\ref{PsiExp}) as
\begin{eqnarray} \label{PsiExp2}
\psi(\mathbf{r}) 
= e^{i\,\theta(\mathbf{r})}
    \,\sqrt{n_0}
    \left(
      e^{i\,\mathbf{k}_0\cdot \mathbf{r}}
        + \frac{1}{2}\frac{\delta n}{n_0}-\frac{1}{8}\frac{\delta n^2}{n_0^2}
    \right)
\, ,
\end{eqnarray}
with $\mathbf{k}_0= \Theta\,\hat{e}_0/L$, $L$ being the length of the
system in the direction of the unit vector $\hat{e}_0$, and $\Theta$ a
small twist angle. For convenience, let us take the order parameter
normalized to unity; the superfluid density - in the direction of
$\hat{e}_0$ - is then defined by the thermodynamic limit of
\begin{eqnarray} 
\rho_s &=& \frac{2\,L^2\,m^2\,N}{\hbar^2\,\Theta^2\,V}
           \left[ 
             F^\Theta(\mu,T)-F^0(\mu,T)
           \right]
\\ \nonumber
      &=& \frac{2\,m^2\,N}{\hbar^2\,k_0^2\,V}
          \left[ 
            F^\Theta(\mu,T)-F^0(\mu,T)
          \right]
\,.
\end{eqnarray}
The substitution (\ref{PsiExp2}) results in the twisted action
\begin{eqnarray} 
\mathcal{S}^\Theta
                &=& \mathcal{S}
                    + \int d\mathbf{r} \, d\tau
                    \frac{\hbar^2}{2\,m}\,
                    \left[
                      k_0^2\, n_0(\mathbf{r},\tau) 
                       - 2\,i\,k_0\, 
                     \delta n(\mathbf{r},\tau)\,\nabla\,i\,\theta(\mathbf{r},\tau)
                   \right]
                \nonumber \\
        &=& \mathcal{S} + k_0\,\int d\tau\,V(\tau) \,,
\end{eqnarray}
where in the second line we ignored higher order terms in the fluctuating 
fields $\delta n$
and $\theta$.  If the system is large enough ($L \gg 1$), then
$V(\tau)$ can be seen as a perturbation; moreover, let us assume that
the symmetry of the problem imposes that the odd terms in $k_0$ 
vanish in $F^\Theta$, since the cases with $k_0$ and $-k_0$ lead to
the same physical situations. The new free energy can be computed
using the linked cluster theorem (see, for example
\cite{Negele:QMP88})
\begin{eqnarray} \label{Ftwist}
F^\Theta &=& F^0 + \sum_{n=1}^\infty \frac{1}{n!}\,\frac{k_0^n}{\beta\,\hbar^n}
                   \int d\tau^1 ... d\tau^n \langle V(\tau^1) ... V(\tau^n) \rangle_{\mathrm{connected}}
                   \nonumber \\
        &=& F^0 + \frac{\hbar^2}{2\,m}\,k_0^2
                - \frac{\hbar^4}{2\,m^2}\,k_0^2\,\frac{1}{\beta}
                \int d\tau \, d\tau'
                \left[                
                  \langle \delta n \, \delta n' \rangle
                  \langle \nabla \theta \, \nabla \theta' \rangle
                  +
                  \langle \delta n \, \nabla \theta' \rangle
                  \langle \nabla \theta \, \delta n' \rangle
                \right]
                + O(k_0^4)   \nonumber \\
        &=& F^0 + \frac{\hbar^2}{2\,m}\, k_0^2 \, \frac{V}{N} 
                                              \,\rho_s + O(k_0^4) 
\,.
\end{eqnarray}
Here, the notation $\langle \ldots \rangle_{\mathrm{connected}}$ stands, 
as usual, for a diagrammatic expansion where only connected diagrams 
are retained. 
The superfluid density can be found in terms of the Green
functions (\ref{Gphipsi}), defining
\begin{eqnarray}
 \tilde{G}(r_1,r_2,r_3,r_4,\tau,\tau')
 &=& 
    G_{11}(r_1,\tau,r_3,\tau')\,G_{00}(r_2,\tau,r_4,\tau')
    \nonumber \\
 &+&
    G_{10}(r_1,\tau,r_4,\tau')\,G_{01}(r_2,\tau,r_3,\tau')\,,
\end{eqnarray}
so that, by partial integration of Eq.\ (\ref{Ftwist}),
\begin{eqnarray} \label{rhos}
\rho_s= \frac{N}{V} 
                &-& \lim_{r_1\rightarrow r}\lim_{r_2\rightarrow r'}
                    \frac{\hbar^2}{m\,\beta} \, \frac{N}{V}\,
                    \int d\mathbf{r} \,d\mathbf{r'} 
                    \,\nabla_1\,\nabla_2
                    \int d\tau \, d\tau'
                      \tilde{G}(r,r',r_1,r_2,\tau,\tau')
                \nonumber \\
                &-& \lim_{r_1\rightarrow r}
                    \frac{\hbar^2}{m\,\beta} \, \frac{N}{V}\,
                    \int d\mathbf{r} \,d\mathbf{r'} 
                    \left( \frac{\nabla \sqrt{n_0(\mathbf{r'})}}{\sqrt{n_0(\mathbf{r'})}}\right) \,                     
                    \,\nabla_1
                    \int d\tau \, d\tau' 
                    \tilde{G}(r,r',r_1,r',\tau,\tau')
                \nonumber \\
                &-& \lim_{r_2\rightarrow r'}
                    \frac{\hbar^2}{m\,\beta} \, \frac{N}{V}\,
                    \int d\mathbf{r} \,d\mathbf{r'} 
                    \left( \frac{\nabla \sqrt{n_0(\mathbf{r})}}{\sqrt{n_0(\mathbf{r})}}\right) \,                    
                    \,\nabla_2
                    \int d\tau \, d\tau' 
                      \tilde{G}(r,r',r,r_2,\tau,\tau')
                \nonumber \\
                &-& 
                    \frac{\hbar^2}{m\,\beta} \, \frac{N}{V}\,
                    \int d\mathbf{r} \,d\mathbf{r'} 
                \left( \frac{\nabla \sqrt{n_0(\mathbf{r})}}{\sqrt{n_0(\mathbf{r})}}\right) \, 
                \left( \frac{\nabla \sqrt{n_0(\mathbf{r'})}}{\sqrt{n_0(\mathbf{r'})}}\right) \,
                    \int d\tau \, d\tau'
                      \tilde{G}(r,r',r,r',\tau,\tau')
\,.
\end{eqnarray}
The evaluation of the averaged operators is done by summing over the
Matsubara frequencies in expressions like
\begin{eqnarray} \label{green_green}
&~&~\frac{1}{\beta}\,\int_0^{\beta} d\tau \,\int_0^{\beta} d\tau'
 G_{\alpha\beta}(\mathbf{r}_1,\tau,\mathbf{r}_2,\tau')
\,
 G_{\gamma\delta}(\mathbf{r}_3,\tau,\mathbf{r}_4,\tau')
\nonumber \\
&=& \frac{1}{\beta^3} \int d\tau \, d\tau'
                    \sum_{m,n}\,\exp(i\,(\omega_n+\omega_m)\,(\tau-\tau'))
                    \,
                    G_{\alpha\beta}(\mathbf{r}_1,\mathbf{r}_2,\omega_n)
                    \,
                    G_{\gamma\delta}(\mathbf{r}_3,\mathbf{r}_4,\omega_m)
\nonumber \\
&=& \frac{1}{\beta}\,
            \sum_n  G_{\alpha\beta}(\mathbf{r}_1,\mathbf{r}_2,\omega_n)
            \,
            G_{\gamma\delta}(\mathbf{r}_3,\mathbf{r}_4,-\omega_n)
\nonumber \\
&=&
\,\sum_{i \ne j}
\Big[
  \frac{N(E_i)+N(E_j)+1}{E_i+E_j}
     \chi_i^\alpha(\mathbf{r}_1)\,{\chi_i^\beta}^\dagger(\mathbf{r}_2)
     \chi_j^\gamma(\mathbf{r}_3)\,{\chi_j^\delta}^\dagger(\mathbf{r}_4)
\nonumber \\
&~&~~~~~~~~~~ - \frac{N(E_i)-N(E_j)}{E_i-E_j}
     \chi_i^\alpha(\mathbf{r}_1)\,{\chi_i^\beta}^\dagger(\mathbf{r}_2)
     \tilde{\chi}_j^\gamma(\mathbf{r}_3)\,\tilde{\chi}_j^\delta{}^\dagger(\mathbf{r}_4)
\nonumber \\
&~&~~~~~~~~~~ - \frac{N(E_i)-N(E_j)}{E_i-E_j} 
     \tilde{\chi}_i^\alpha(\mathbf{r}_1)\,\tilde{\chi}_i^\beta(\mathbf{r}_2)
     \chi_j^\gamma(\mathbf{r}_3)\,\chi_j^\delta{}^\dagger(\mathbf{r}_4)
\nonumber \\
&~&~~~~~~~~~~ +
  \frac{N(E_i)+N(E_j)+1}{E_i+E_j} 
     \tilde{\chi}_i^\alpha(\mathbf{r}_1)\,\tilde{\chi}_i^\beta{}^\dagger(\mathbf{r}_2)
     \tilde{\chi}_j^\gamma(\mathbf{r}_3)\,\tilde{\chi}_j^\delta{}^\dagger(\mathbf{r}_4)
\Big]
\nonumber \\
&+&
\sum_i 
\frac{2\,N(E_i)+1}{2\,E_i} 
\,
\left[
     \chi_i^\alpha(\mathbf{r}_1)\,{\chi_i^\beta}^\dagger(\mathbf{r}_2)
     \chi_i^\gamma(\mathbf{r}_3)\,{\chi_i^\delta}^\dagger(\mathbf{r}_4)
+
     \tilde{\chi}_i^\alpha(\mathbf{r}_1)\,\tilde{\chi}_i^\beta{}^\dagger(\mathbf{r}_2)
     \tilde{\chi}_i^\gamma(\mathbf{r}_3)\,\tilde{\chi}_i^\delta{}^\dagger(\mathbf{r}_4)
\right]
\nonumber \\
&+&
\beta
\sum_i 
N(E_i)\,( N(E_i) + 1 )
\,
\left[
     \tilde{\chi}_i^\alpha(\mathbf{r}_1)\,\tilde{\chi}_i^\beta{}^\dagger(\mathbf{r}_2)
     \chi_i^\gamma(\mathbf{r}_3)\,{\chi_i^\delta}^\dagger(\mathbf{r}_4)
+
     \chi_i^\alpha(\mathbf{r}_1)\,{\chi_i^\beta}^\dagger(\mathbf{r}_2)
     \tilde{\chi}_i^\gamma(\mathbf{r}_3)\,\tilde{\chi}_i^\delta{}^\dagger(\mathbf{r}_4)
\right]
\,,
\end{eqnarray}
with $\chi_i$ and $\tilde{\chi_i}$ defined in Eq. (\ref{Chi}) and
(\ref{ChiTilde}). In particular, for the uniform case, all the terms
proportional to $\nabla\sqrt{n_0}$ vanish; moreover, since the
expressions for the Bogoliubov functions are known analytically, 
it is possible to
see that only the last term of (\ref{green_green}) survives, each term
of the sum between square parenthesis giving a contribution of
$k^2/2$. In one dimension one obtains 
\begin{eqnarray}
\rho_s= \frac{N}{V} - \frac{\hbar^2\,N}{m\,V} \sum_{k\ne0} k^2\, \frac{\partial N(E_k)}{\partial E_k}
\, ,
\end{eqnarray}
which is the well known Landau result \cite{LandauVol9}.

\section{Bose glass transition} \label{sec:boseglass} 
\twocolumngrid

We now apply our expressions for the correlation function and the
superfluid density to a system at $T=0 $. The external potential is
quasiperiodic, obtained as a sum of two potentials whose periods are
incommensurate with each other, 
\begin{equation} \label{extern}
U(x)= V_1\,\cos\left(\frac{2\,\pi}{d}\,x\right)
    + V_2\,\cos\left( \frac{2\,\pi}{\lambda d}\,x\right)\,.
\end{equation}
For our specific realization we have chosen $\lambda$ to be the golden
ratio, approximated as a fraction of two consecutive Fibonacci
numbers. In the following we shall consider two systems with length
$L=377\,d$ and $L=89\,d$; the value of $\lambda$ is, respectively,
$\lambda=377/233$ and $\lambda=89/55$.

For $V_2=0$, upon increasing the value of $V_1$, the system ceases to
be superfluid and enters a Mott insulating phase \cite{fisher1989,
  jaksch1998, greiner2002}.  As stated in the introduction, this phase
cannot be seen using the Bogoliubov ansatz we employ for the
excitations, and our method should not be applied when the physics of
the system is affected by the presence of the Mott phase.  However,
when the second lattice is turned on a new quantum phase, the Bose
glass, becomes possible. According to Ref. \cite{fisher1989}, when
$V_2$ is greater than the interaction energy, the Mott phase
disappears from the phase diagram of the gas, and the only insulating
phase is the glass phase. We believe that in this situation of strong
disorder our approach can be quantitatively correct, and for 
definiteness we have chosen the weights of the two potentials to be
equal ($V_1=V_2=V$).

Taking $d$ as the unit of length, the Gross-Pitaevskii equation can be
rewritten as
\begin{equation} \label{GP2}
\left[
  -\frac{1}{2}\frac{d^2}{dx^2}
  + g\,N\,|\Phi(x)|^2
  + U(x)
\right]
\,\Phi(x)
=
\mu\,\Phi(x)
\,,
\end{equation}
where the energy is measured in units of $E_d=\hbar^2/m\,d^2$, and the
norm of the quasicondensate wavefunction 
$\Phi$ is set to $1$. For each value of the parameters $V$ and
$g$, we have found the ground state of Eq.~(\ref{GP2}) using an
imaginary time evolution with the split-operator method, along with
periodic boundary conditions. Using the ground-state wavefunction the
Bogoliubov excitations are found by means of a direct diagonalization
of the Bogoliubov equations (\ref{BdG}). In all our
calculations the Laplacian term is represented using the Fourier
transform, while the derivative in the expression for superfluid density
(\ref{rhos}) is approximated using a finite difference expression.

\begin{figure}[tbp]
\centering
\includegraphics[width=3.7in,clip]{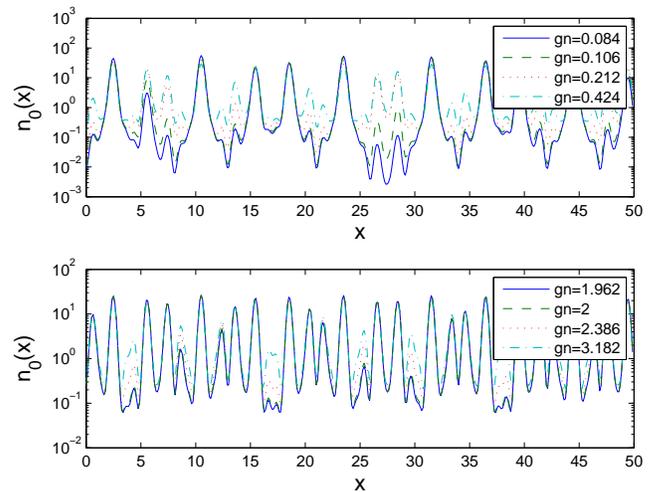}
\caption{(Color online) Density profiles for some values of $g\,n$ for
  $V=5\,E_d$ (top panel) and $V=9\,E_d$ (bottom panel); $g\,n$ is
  expressed in units of $E_d=\hbar^2/m\,d^2$ and $x$ in units of $d$.
  Upon decreasing the interaction strength the condensate breaks up in
  sets of spikes with little overlap.}
\label{dens}
\end{figure}

The Bogoliubov analisis is expected to be relevant for values of the
interparticle interaction $g/n \ll 1$. This means that our work is
realistic for 
\begin{equation}
g\,n \ll n^2\,,
\end{equation}
i.e., the approximation we are using starts to be meaningful for
systems that contain a few particles per site.

We first solve for the ground state in a system with $L=377\,d$, using
a numerical grid of $3016$ points.  Choosing a mean density
$n=N/L=4$, we show in Fig.~(\ref{dens}) $n_0(x) = N\,|\Phi(x)|^2$ for
the two cases $V=5\,E_d$ and $V=9\,E_d$. Upon decreasing $g$, the
condensate density develops dips that become more and more pronounced.
Indeed, in some regions the condensate seems to be broken up into
several pieces that hardly overlap.  As we shall see, this has
consequences for the behavior of the correlation function.

\begin{figure}[tbp]
\centering
\includegraphics[width=3.7in,clip]{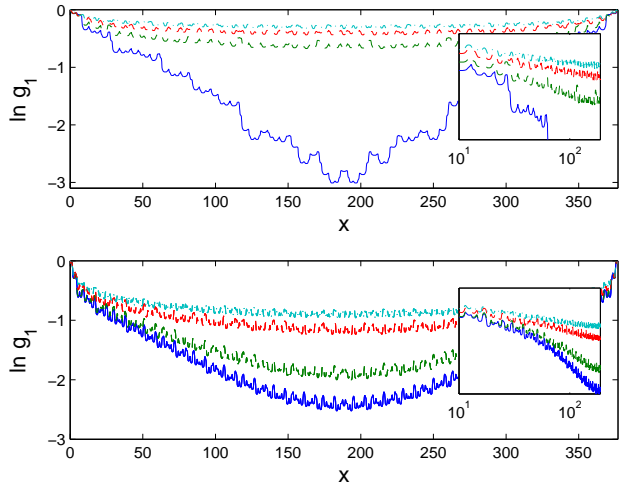}
\caption{(Color online) Exponent of the correlation function
  (\ref{g1}) for the same values of $g\,n$ as in Fig \ref{dens}, for
  $V=5\,E_d$ (top panel) and $V=9\,E_d$ (bottom panel); $g\,n$ is
  expressed in units of $E_d=\hbar^2/m\,d^2$ and $x$ in units of $d$.
  The insets shows the same plot with a log scale on the horizontal
  axis. In the main picture it is seen that for lower $g\,n$ the
  values of $\ln\,g_1$ (\ref{g1}) follow a linear behavior.  For
  higher $g\,n$ the plot appears to depend logarithmically on $x$.}
\label{corr}
\end{figure}

\begin{figure}[tbp]
\centering
\includegraphics[width=3.7in,clip]{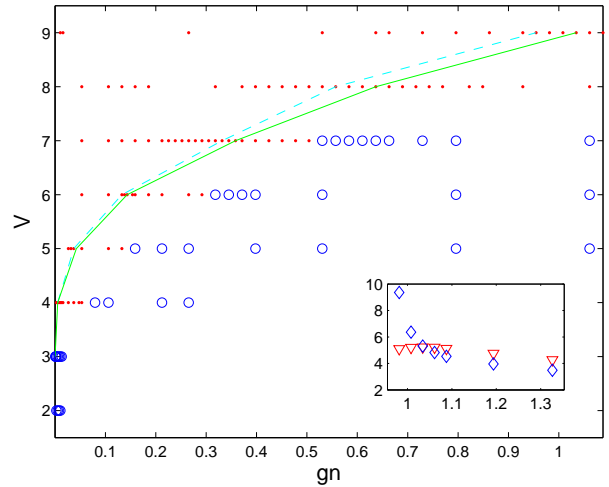}
\caption{(Color online) Phase diagram of the Bose glass transition. The open circles
  correspond to a state with non-vanishing superfluid component, while
  for the red dots the system has only a normal part. The filled line
  indicates when the linear interpolation of $\ln\,g_1$ (Eq.\ (\ref{g1})) 
  gives
  a smaller error than a logarithmic fit, i.e., at the left side of
  the filled line the correlation function decays exponentially.  The
  dashed line 
  indicates the point at which the smallest excitation energies 
  cannot be resolved numerically. 
  The inset shows the sum of the residuals of the fits (in arbitrary
  units) as a function of $g\,n$, for $V=9\,E_d$: the triangles refer
  to a linear regression, the diamonds to a logarithmic interpolation.
  $V$ and $g\,n$ are expressed in units of $E_d=\hbar^2/m\,d^2$.}
\label{inter}
\end{figure}

Using the same mean density $n=N/L=4$, Fig.~(\ref{corr}) plots
the exponent of the correlation function
\begin{equation}\label{g1}
  \ln\, g_1(0,x)= \ln\,\langle\psi(0)\,\psi(x)\,\rangle - 
  1/2\,\ln\,n_0(0) n_0(x)\,.
\end{equation}
While for higher $g\,n$ the values of $\ln\,g_1$ follow a logarithmic
behavior (insets of Fig.~(\ref{corr}) ), leading to a power law decay
in the correlation function, for lower $g\,n$ the function shows a
linear fall, therefore giving an exponential decay. Since our system
is finite the transition to a different decay behavior is gradual, but
as a quantitative measure we record the point where a linear fit for
$\ln\,g_1$ gives a smaller sum of the residuals than a logarithmic
fit.  The full line in Fig.~(\ref{inter}) shows the critical value of
$V$ as a function of the interaction strength, according to an
interpolation of $\ln\,g_1$ between $x=10\,d$ and $x=150\,d$.

The change in the decay behavior becomes evident at the values of $g$
when the quasicondensate seems to be broken up into a set of spikes
with little overlap. In this situation the lowest energy excited modes
have a phase flip character, since it costs little energy to change
the relative phase of two barely overlapping zones; the energy of the
lowest excitation decreases upon decreasing $g$.  Eventually, this
value becomes so low that reaches the limit of our numerical precision
($\sim 10^{-15}\,E_d$).  At this point, we may say that the condensate
has become completely disjoint.  This boundary is indicated by the
dashed line in Fig.~(\ref{inter}).  It is seen that it occurs close to
the Bose glass transition, but it does not coincide with the
transition found using the other criteria.  In the lower left corner
of the phase diagram, for small values of the interparticle potential,
detecting the transition can be numerically challenging, since at the
point where the correlation function changes its behavior the
excitation spectrum is already at the boundaries of the numerical
precision.

Figure (\ref{inter}) also shows when the superfluid density, as given
by Eq.~(\ref{rhos}), drops to zero (red points). Apparently, the
superfluid formula systematically overestimates the critical
interaction strength for the transition to an exponential decay.
Indeed, Eq.~(\ref{rhos}) gives the superfluid density only in the
thermodynamic limit and it might be that the finite size of the system 
is part of the explanation for the disagreement. 
However, we believe there is another reason for 
this discrepancy: The way we applied the phase twist to the condensate
part in Eq.~(\ref{PsiExp2}). For a non-uniform system, a phase twist
in the boundary conditions should give a velocity that depends on the
coordinate, as a function of the condensate density in each point. The
corrections to (\ref{rhos}) should be more relevant for small $g$,
when the variations in density are bigger.

The interpretation of the phase diagram in Fig. (\ref{inter}) is the
following: at the left side of the line there is an exponentially
decaying correlation function, along with a vanishing superfluidity;
we believe that at this point our Bogoliubov scheme detects the Bose
glass phase, as described previously in Sec.\ \ref{sec:intro}.  The
Bose glass phase is seen to disappear completely below a finite value
for the disorder strength $V$ (approximately $V=3$), in contrast to
the findings of Ref.\ \cite{fontanesi2009}. This is the main
difference with the results of Fontanesi \emph{et al.}. Besides finite
size effects, this behavior seems to be characteristic of the
quasiperiodic potential. As stated in Sec. \ref{sec:intro}, for
vanishing interaction the system is localized only when the external
potential strength is higher than a critical value. When a repulsive
interaction is considered, this property survives in the
glassy phase.

\begin{figure}[tbp]
\centering
\includegraphics[width=3.7in,clip]{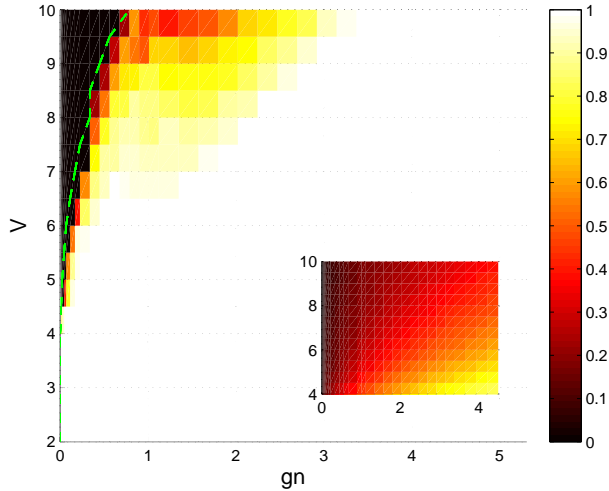}
\caption{(Color online) Surface plot of the superfluid density (\ref{rhos})
  (normalized to the total density $n$) as a function of $V$ and
  $g\,n$, for $L=89\,d$. The dashed line is the boundary that we
  estimate for the validity of our numerical calculations (see
  text). The inset shows the inverse participation ratio $P$
  (\ref{part_ratio}) of the quasicondensate wavefunction.  Both $V$
  and $g\,n$ are measured in units of $E_d=\hbar^2/m\,d^2$.}
\label{superfluid}
\end{figure}

In Fig.~\ref{superfluid}, we show the behavior of $\rho_s$ in an
enlarged phase diagram.  Because of the computational burden, this
phase diagram was calculated for a smaller system with $L=89\,d$ and a
grid of $1424$ points, and therefore the Bose glass boundary is
slightly shifted compared to that in Fig.~\ref{inter}.  Monte Carlo
simulations \cite{Scalettar1991, Prokofev1998} have previously shown a
reentrant phase transition upon increasing $g$. While we do not see
this re-entrance, the behavior of the superfluidity is suggestive, as
it shows a non-monotonous dependence of the normal phase with respect
to the interaction and the external potential strength. Indeed, for a
quasiperiodic potential the paper Ref. \cite{roux2008} shows that the
re-entrant phase is limited to a small portion of the phase space, and
it is compatible with a superfluid phase that extends to higher
interaction strengths. Within the limit of the Bogoliubov
approximation, our work agrees with this conclusion.

\begin{figure}[tbp]
\centering
\includegraphics[width=3.7in,clip]{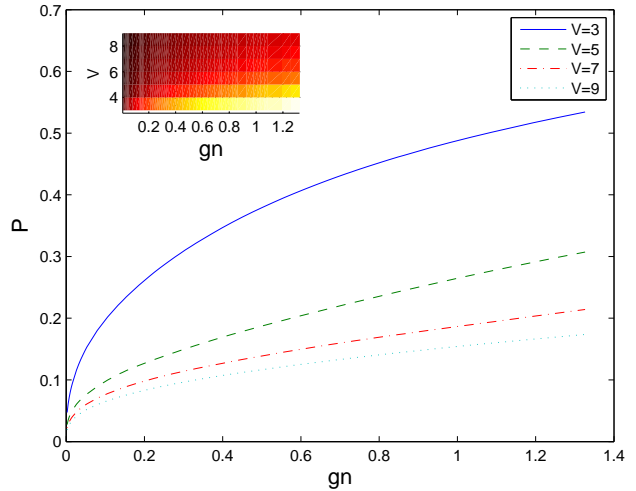}
\caption{(Color online) Inverse participation ratio $P$ (\ref{part_ratio}) for
  $L=377\,d$ as a function of $g\,n$, for some values of the external
  potential strength $V$. The inset shows a surface plot of $P$ as a
  function of $V$ and $g\,n$. All the energies are measured in units
  of $E_d=\hbar^2/m\,d^2$.}
\label{p_ratio}
\end{figure}

Finally, Fig (\ref{p_ratio}) and the inset of Fig. (\ref{superfluid})
shows the inverse participation ratio of the quasicondensate
wavefunction $\Phi$,
\begin{equation} \label{part_ratio}
P=\frac{1}{L} \frac{(\int dx |\Phi(x)|^2)^2}{\int dx |\Phi(x)|^4} \,,
\end{equation}
for both of the cases $L=377\,d$ and $L=89\,d$. Since the computation
of $P$ is much faster than the diagonalization of the Bogoliubov
equations, Fig. (\ref{p_ratio}) presents the behavior of the system
for a finer resolution in the values of $g\,n$ and $V$ than Fig.
(\ref{inter}). As we can see, the plots do not show any distinct
transition. The glassy phase does not appear to be related to a
radical change in the density profile.  Moreover, as stated in Sec.
\ref{sec:intro}, a new phase, the Lifshits glass, has been conjectured
to exist for weak interparticle potential, where the peaks in the
density profile do not overlap. We argue that such a phase should have
a distinctive signal in the quantity $P$.  However, the plots of the
inverse participation ratio do not show only a smooth decrease.  We
conclude that - within our approximation - the Lifshits glass does not
seem to exist as a phase in its own right, but only as the limit of
vanishing interaction.

\section{Conclusions} \label{sec:conclusion}

Using a path integral formalism we have obtained -- within the
Bogoliubov approximation -- expressions for the correlation function
and the superfluid density in a Bose gas for an arbitrary external
potential, at zero or finite temperature.  Applying these expressions
to a one-dimensional system at $T=0$, a quasi-periodic external
potential was seen to cause a transition to a phase where the
correlation decays exponentially and the superfluid density vanishes.
We believe this is the Bose glass phase. A comparison with the
experiment of Fallani \emph{et at.}  \cite{fallani2007} is not
straightforward, because they employed a quasiperiodic potential with
different relative weights [i.e., $V_1\neq V_2$ in Eq.
(\ref{extern})], and the proximity to the Mott insulating phase in the
experiment makes the Bogoliubov approximation questionable.  However,
they spotted signs of a Bose glass transition when the peak height of
the external potential varies between $2V\approx16\,E_d$ and
$2V\approx18\,E_d$, with $g\,n\approx 1\,E_d$, which is where the
transition takes place in our study for the bigger system
($L=377\,d)$.

We have not found a perfect match between the results for the
superfluid density and the correlation behavior for the location of
the Bose glass transition. We would like to point out that the formula
for the superfluid density (\ref{rhos}) assumes that the velocity
$v_s$ of the condensate does not depend on the position, and in this
sense it is an approximation even within the Bogoliubov approach. For
this reason we plan to improve the expression for $\rho_s$ in the
future.

Finally, we notice that our expression for the correlation function
and the superfluid density are valid also at a finite temperature, and
they can be used to study the Bose glass transition for $T\ne0$. This
relevant issue will be considered in a future work.

\begin{acknowledgments}
  This research was conducted using the resources of the High
  Performance Computing Center North (HPC2N), and financially
  supported by the Swedish Research Council, Vetenskapsr{\aa}det. We
  thank Magnus Johansson and Andrei Shelankov for enlightening
  discussions.
\end{acknowledgments}



\end{document}